# Superoxide Reductase from *Desulfoarculus baarsii* : Reaction Mechanism and Role of Glutamate 47 and Lysine 48 in Catalysis†

Murielle Lombard ‡, Chantal Houée-Levin §, Danièle Touati ⊠

Marc Fontecave ‡, and Vincent Nivière ‡*

‡ Laboratoire de Chimie et Biochimie des Centres Redox Biologiques, DBMS-CEA/CNRS/Université Joseph Fourier, 17 Avenue des Martyrs, 38054 Grenoble, Cedex 9, France. § Laboratoire de Chimie Physique, CNRS/Université Paris-Sud, Bâtiment 350, Centre Universitaire F-91405 Orsay Cedex, France. ⊠Institut Jacques Monod, CNRS/Universités Paris 6 et Paris 7, 2 place Jussieu, 75251 Paris Cedex 05, France

Running Title : Mechanism of Superoxide Reductase by Pulse Radiolysis

* To whom correspondence should be addressed. Tel. : 33-4-76-88-91-09;  Fax : 33-4-76-88-91-24; E-mail : vniviere@cea.fr

ABBREVIATIONS

[1] The abbreviations used are : SOD, superoxide dismutase; CuZn SOD, copper-zinc superoxide dismutase; SOR, superoxide reductase; Dfx, desulfoferrodoxin; EPR, electron paramagnetic resonance; PAGE, polyacrylamide gel electrophoresis; PCR, polymerase chain reaction.

ABSTRACT

Superoxide reductase (SOR) is a small metalloenzyme which catalyzes reduction of $O_2^-$ to $H_2O_2$ and thus provides an antioxidant mechanism against superoxide radicals. Its active site contains an unusual mononuclear ferrous center, which is very efficient during electron transfer to $O_2^-$ [Lombard, M., Fontecave, M., Touati, D., and Nivière, V. (2000) Journal of Biological Chemistry 275, 115-121]. The reaction of the enzyme from *Desulfoarculus baarsii* with superoxide was studied by pulse radiolysis methods. The first step is an extremely fast bi-molecular reaction of superoxide reductase with superoxide, with a rate constant of $(1.1\pm0.3)\times10^9$ $M^{-1}$ $s^{-1}$. A first intermediate is formed which is converted to a second one at a much slower rate constant of $500\pm50$ $s^{-1}$. Decay of the second intermediate occurs with a rate constant of $25\pm5$ $s^{-1}$. These intermediates are suggested to be iron-superoxide and iron-peroxide species. Furthermore, the role of glutamate 47 and lysine 48, which are the closest charged residues to the vacant sixth iron coordination site, has been investigated by site-directed mutagenesis. Mutation of glutamate 47 into alanine has no effect on the rates of the reaction. On the contrary, mutation of lysine 48 into an isoleucine led to a 20-30-fold decrease of the rate constant of the bi-molecular reaction, suggesting that lysine 48 plays an important role during guiding and binding of superoxide to the iron center II. In addition, we report that expression of the lysine 48 *sor* mutant gene hardly restored to a superoxide dismutase-deficient *Escherichia coli* mutant the ability to grow under aerobic conditions.

Superoxide radical ($O_2^-$) is the univalent reduction product of molecular oxygen. It belongs to the group of the so-called toxic oxygen derivatives, which also include hydrogen peroxide and hydroxyl radicals (*1*). For years the only enzymatic system known to catalyze the elimination of superoxide was the superoxide dismutase (SOD) [1], which catalyzes dismutation of superoxide radical anions to hydrogen peroxide and molecular oxygen (*2*) :

$$O_2^- + O_2^- + 2 H^+ \rightarrow H_2O_2 + O_2$$

Very recently, a new concept in the field of the mechanisms of cellular defense against superoxide has emerged (*3, 4*). It was discovered that elimination of $O_2^-$ could also occur by reduction, a reaction catalyzed by an enzyme thus named superoxide reductase (SOR) :

$$SOR_{red} + O_2^- + 2 H^+ \rightarrow H_2O_2 + SOR_{ox}$$

$$SOR_{ox} + 1e^- \rightarrow SOR_{red}$$

Two classes of SOR have been described. The first one is a small protein found in anaerobic sulfate-reducing and microaerophilic bacteria, initially called desulfoferrodoxin (Dfx). It is a homodimer of 2x14kDa, which has been extensively studied (*3, 5-8*), and the protein from *Desulfovibrio desulfuricans* has been structurally characterized (*9*). The monomer is organized in two protein domains. The N-terminal domain has a fold similar to that of desulforedoxin (*10*) and contains a mononuclear ferric iron, center I, coordinated by four cysteines in a distorted rubredoxin-type center (*9*). In the SOR from the microaerophilic bacteria *Treponema pallidum* only one of these cysteines is conserved and the N-terminal domain does not chelate iron (*8*). The second domain of SOR contains a different mononuclear iron center, center II, consisting of an oxygen-stable ferrous iron with square-pyramidal coordination to four nitrogens from histidines as equatorial ligands and one sulfur from a cysteine as the axial ligand (*9*). The midpoint redox potentials have been reported to be 2-4 mV for center I, and 90-240 mV for center II (*6, 7*).

It has been demonstrated that the iron center II of SORs from *Desulfoarculus baarsii* and *Treponema pallidum* is the active site as it reduces superoxide very efficiently, without significant SOD activity (*3, 8*). A second order rate constant for the reaction of iron center II with superoxide has been determined to be $0.7$-$1 \times 10^9$ $M^{-1}$ $s^{-1}$ (*3, 8*), a value comparable to the value of the rate constant determined for SODs (*11, 12*). In addition, the active site of SOR is specific for $O_2^-$ since the reduced iron center II is not oxidized by $O_2$ and only very slowly by $H_2O_2$ (*3, 8*). How a reduced iron center II is regenerated for a second cycle with $O_2^-$ has not yet been clearly demonstrated and the electron donor to SOR remains to be identified.

Although SOR is not naturally present in *E. coli* it was demonstrated that expression of SORs from *D. baarsii* and *T. pallidum* in *E. coli* could totally replace the SOD enzymes to overcome a superoxide stress (*8, 13*). That SOR was an efficient antioxidant protein also in sulfate-reducing bacteria, its natural host, was further shown from the observation that a *Desulfovibrio vulgaris Hildenborough* mutant strain lacking the *sor* gene was very oxygen-sensitive during transient exposure to microaerophilic conditions (*14*).

A second class of SOR has been characterized from the anaerobic archaeon, *Pyrococcus furiosus* (*4*). The homotetrameric protein presented strong homologies to neelaredoxin (Nlr), a small protein containing a single mononuclear center, earlier characterized from sulfate-reducing bacteria (*15*). Very recently, the three-dimensional structure of the *P. furiosus* SOR has been determined at high resolution (*16*). The overall protein fold is similar to that of the iron center II domain of SOR from *D. desulfuricans* and the structure of the unique mononuclear iron center is similar to that of the iron center II. Obviously, there is no rubredoxin-like center I domain in SOR from *P. furiosus*. This again is consistent with the notion that center II is the active center. Interestingly, the crystal structure of SOR from *P. furiosus* suggested that the sixth iron coordination site of center II might be occupied by a glutamate, strongly conserved among all the primary known sequences of SORs, only in the oxidized state. In the reduced form instead, this site is vacant or occupied by solvent molecules and is thus potentially accessible to $O_2^-$ (Scheme 1). However, whether this observation was an artifact of crystallization or the manifestation of an important feature of the enzymatic mechanism of SOR needs further investigation. In addition, the presence of a strictly conserved lysine residue, close to the vacant sixth iron-coordination site, suggests that this position may play a role in facilitating the binding of superoxide to the iron center (*9, 16*) (Scheme 1).

In this work, in order to learn more about the mechanism of action of SOR, we have studied the reaction of the enzyme from *Desulfoarculus baarsii* with superoxide by the mean of pulse radiolysis and site-directed mutagenesis. Pulse radiolysis, which allows specific and quantitative production of superoxide, when coupled with spectrophotometric detection, is particularly adapted to study such a reaction. We have found that during reaction of the iron center II with superoxide, at least two successive reaction intermediates are formed. Furthermore, the role of Glu47 and Lys48 which correspond to the Glu and Lys residues discussed above (Scheme 1) has been investigated by site-directed mutagenesis and pulse radiolysis. Only Lys48 seems to play an important role for catalysis.

MATERIALS AND METHODS

*Materials*. 1-2 mM $KO_2$ stock solutions were prepared in anhydrous $Me_2SO$ as described in (*3*). Xanthine oxidase Grade IV from milk (0.19 unit/mg), catalase from *Aspergillus niger* (6600 units/mg), cytochrome *c* from bovine heart, CuZn SOD from bovine erythrocytes (6000 units/mg) were from Sigma. For pulse radiolysis experiments, sodium formate and Tris base were of the highest quality available (Prolabo Normatom or Merck Suprapure). Oxygen was delivered by ALPHA GAZ. Its purity is higher than 99.99%. Water was purified using an Elga Maxima system (resistivity 18.2 MΩ).

*Bacterial Strain and PCR-Based Site-directed Mutagenesis*. *E. coli* strain QC 2375 (*sodA sodB recA*) was previously described (*13*). SOR mutants were generated from the plasmid pMJ25, in which the *sor* gene from *Desulfoarculus baarsii* was cloned (*13*). Six primers were designed and used for PCR-based site-directed mutagenesis to create the two mutants, Glu47Ala SOR and Lys48Ile SOR. Primer 1 (5'A<u>GAATTC</u>GAGCTCAAGTGCG), and primer 2 (5'AA<u>AAGCTT</u>AAATCTCCAGAGAGTGAAC), were used to generate the entire Dfx coding region (515 bp) with an *Eco*R1 and a *Hind*III restriction site (underlined), respectively. Primers 3 (5'GTGCTT<u>GG</u>CCTTGGCCGCATC for Glu47Ala and 5'CGGCACGTG<u>GA</u>TTTCCTTGGC for Lys48Ile) and primers 4 (5'GCCAAGG<u>CC</u>AAGCACGTG for Glu47Ala and 5'GCCAAGGAAA<u>TC</u>CACGTGCCG for Lys48Ile) contained the mutation of interest (underlined). For each mutation, two separate PCR reactions were used to amplify the 5' portion (primers 1 and 3) and 3' portion (primers 2 and 4) of the *sor* DNA from pMJ25. The PCR products were purified and used as a template DNA for a second round of PCR using primers 1 and 2. The resulting Glu47Ala and Lys48Ile *sor* PCR products were double digested with *Eco*RI and *Hind*III and inserted into the corresponding sites of the expression vector pJF119EH, under *ptac* promotor control (*13*). The two resulting plasmids, pMLE47A and pMLK48I were transformed in *E. coli* DH5α. The constructs and the mutations were verified by DNA sequencing.

*Purification of SORs*. The wild-type and mutant SOR proteins were purified according to the same procedure. *E. coli* QC 2375/pMJ25 or DH5α/pMLE47A or DH5α/pMLK48I cells were grown aerobically at 37 °C in M9 minimal medium complemented with 0.4 % glucose, 2 µg/ml thiamin, 1 mg/ml casamino acid, 1 mM $FeSO_4\ 7H_2O$ and 100 µg/ml ampicilin (8 x 1000 ml in a 2-liter Erlenmeyer flask). 2 mM IPTG were added at OD 600 nm of 0.3. At OD 600 nm of about 2.0, cells were chilled and collected by centrifugation. All the following operations were carried out at 4 °C and pH 7.6. The cell pellet was suspended in 3 volumes (w/v) of 0.1 M Tris/HCl and sonicated. After ultracentrifugation at 45,000 rpm during 90 min in a Beckman 50.2 Ti rotor, the supernatant was treated with streptomycin sulfate (final concentration 2% w/v), and after centrifugation was then precipitated with ammonium sulfate (final concentration 80 % w/v). The pellet was dissolved in 12 ml of 25 mM Tris/HCl and loaded onto an ACA 54 column (360 ml) equilibrated with the same buffer. A low molecular weight fraction was collected. Protein fractions of 20-30 mg were further chromatographed onto an anion exchange column Uno Q-6 (Bio-Rad), equilibrated with 10 mM Tris/HCl, using a Bio-Rad Biologic system. A linear NaCl gradient was applied (0-0.15 M NaCl) with the same buffer, with a flow rate of 2 ml.min$^{-1}$ during 50 min. Fractions (about 7 mg) were eluted at 70 mM, 50 mM and 80 mM NaCl for SOR wild-type, SOR Glu47Ala and SOR Lys48Ile, respectively. In all cases, the fractions exhibited a $A_{280nm}/A_{503nm}$ ratio of 4.8-4.9. At this stage, the three SOR proteins appeared to be homogeneous, as shown by SDS-PAGE analysis (15% acrylamide). Protein concentration was determined using the Bio-Rad protein assay reagent (*17*). Protein-bound iron was determined by atomic absorption spectroscopy. Mass spectra were obtained on a Perkin-Elmer Sciex API III+ triple quadrupole mass spectrometer equipped with a nebulizer-assisted electrospray source operating at atmospheric pressure. EPR measurements were made on a Bruker EMX 081 spectrophotometer equipped with an Oxford Instrument continuous flow cryostat.

*Steady-State Measurement of the Kinetic Parameters of the oxidation of SOR by $O_2^-$*. The kinetics of the oxidation of SOR by $O_2^-$, generated by the xanthine-xanthine oxidase system, was followed spectrophotometrically, in the absence or in the presence of different amounts of CuZn SOD, as reported previously (*3*). At the concentration of SOD which decreases by 50% the rate of oxidation of SOR, one can write : $k_{SOD}$ [SOD] = $k_{SOR}$ [SOR] (Eq. 1) where $k_{SOR}$ and $k_{SOD}$ are the second-order rate constants of the reaction of SOR and SOD with $O_2^{·-}$, respectively. Taking into account the known second-order rate constant of the reaction of $O_2^{·-}$ with CuZn SOD at low [$O_2^{·-}$], $2 \times 10^9$ $M^{-1}$ $s^{-1}$ (*11*), the second order rate constant of the oxidation of SOR by $O_2^{·-}$, $k_{SOR}$, was calculated using Eq. 1 (*3*).

*Pulse Radiolysis Experiments*. Free radicals were generated by the application into an aqueous solution of a 200 ns pulse of high energy electrons, ca. 4 MeV from a linear accelerator located at the Curie Institute, Orsay France (*18*). The doses per pulse (2-15 Gy) were calibrated from the absorption of the thiocyanate radical $SCN^{·-}$ obtained by radiolysis of thiocyanate ion solution in $N_2O$-saturated phosphate buffer ([$SCN^-$] = $10^{-2}$ M, 10 mM phosphate, pH 7, G($SCN^{·-}$)= 0.55 µmol. $J^{-1}$, 472 nm, ε = 7580 $M^{-1}$ $cm^{-1}$) (*19*). Superoxide radicals were generated during scavenging of radiolytically generated $HO^·$ free radicals by 100 mM formate, in $O_2$ saturated solution :

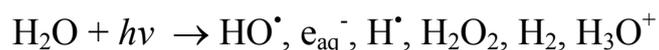
$$H_2O + h\nu \rightarrow HO^·, e_{aq}^-, H^·, H_2O_2, H_2, H_3O^+$$

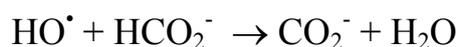
$$HO^· + HCO_2^- \rightarrow CO_2^{·-} + H_2O$$

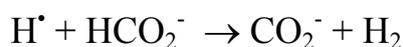
$$H^· + HCO_2^- \rightarrow CO_2^{·-} + H_2$$

$$CO_2^- + O_2 \rightarrow CO_2 + O_2^{\bullet-}$$

$$e_{aq}^- + O_2 \rightarrow O_2^{\bullet-}$$

$$H^{\bullet} + O_2 \rightarrow HO_2^{\bullet}$$

$$HO_2^{\bullet} \rightarrow H^+ + O_2^{\bullet-} \quad (pKa = 4.7)$$

This radical is obtained in pure form in less than one microsecond (*20*). Unless otherwise stated, samples to be irradiated were made up in 10 mM Tris/HCl buffer, pH 7.6, 100 mM sodium formate, and saturated with pure $O_2$. The doses per pulse were ca. 5 Gy ($[O_2^-] \approx 3$ μM). The $O_2^-$ concentration and self-decay were also assayed from its own absorbance at 271 nm (molar extinction coefficient of 1400 $M^{-1}$ $cm^{-1}$) and compared to literature data (*21*). The protein concentration was always between 70 and 150 μM. The reaction was followed spectrophotometrically between 450 and 750 nm, in a 2 cm path length cuvette designed for pulse radiolysis experiments. Time-dependent difference absorbances were recorded on a digital oscilloscope from two different experiments at two different time scales. Kinetics were always analyzed at different wavelengths. The traces could always be fitted according to a first order kinetic law. For the reaction of superoxide with SOR or any mutant, apparent rate constants were always proportional to protein concentration (between 70 and 150μM), hence the rate constants were equal to the slopes of straight lines $k_{app}$ versus [protein]. The spectra of the intermediates were reconstructed from the absorbances at maxima in the wavelength range investigated.

RESULTS

*Expression and Purification of the SOR mutants*. The level of expression in *E. coli* of the Glu47Ala and the Lys48Ile mutant forms of SOR from *D. baarsii* was similar to that of the recombinant wild-type enzyme (*3*). The two mutants were purified by following a similar protocol to that for the wild-type SOR (*3*) and were both obtained in homogeneous form, as shown by SDS-PAGE electrophoresis. Electrospray mass spectrometry analysis showed a major species at 13,968 Da and at 14,011 Da for the Glu47Ala and the Lys48Ile mutant forms of SOR, respectively, corresponding to the molecular masses expected from the amino acid sequences, without the N-terminal Met residue (*13*). The iron content of both SOR mutants was determined by atomic absorption spectroscopy. Values of 2.0 and 1.9 iron atoms per polypeptide chain were found for the Glu47Ala and the Lys48Ile mutants, respectively. No evidence for the presence of zinc or manganese atoms was found in both mutant proteins (data not shown).

*Spectral Properties of Glu47Ala and Lys48Ile mutants*. As shown in Fig. 1, the UV-visible spectra of the as-isolated Glu47Ala and the Lys48Ile mutants exhibited the usual absorption bands at 370 nm and 503 nm, contributed by the ferric iron from center I (*5, 7*). The ratio $A_{280\,nm}/A_{503\,nm}$ was about 4.8 for both mutant proteins and a value of the molar extinction coefficient at 503 nm was determined to be 4,400 M$^{-1}$ cm$^{-1}$ in both cases, similar to that reported for the wild-type SOR (*3*). When treated with successive additions of stoichiometric amounts of O$_2^-$, both mutants exhibited an increase of the absorbances in the 500-700 nm range, reflecting oxidation of iron of center II (*3, 7*) (Fig. 1). Full oxidation was achieved after addition of 4 to 8 molar excess of O$_2^-$, as in the case of the wild-type protein (*3*).

The oxidized Lys48Ile mutant (Fig. 1B) exhibited a spectrum comparable to that of the oxidized wild-type protein (Fig. 1A) (so-called gray form), with absorption bands contributed by the ferric irons from both centers I and II (*7*). However, the band specific for center II was centered at 635 nm, a value slightly different from that observed in the case of the wild-type protein ($\lambda_{max}$ = 644 nm) (*3*). The molar extinction coefficient at 635 nm was found to be 1,400 mM$^{-1}$ cm$^{-1}$, a slightly lower value with respect to that for the wild-type SOR at 644 nm (1,900 mM$^{-1}$ cm$^{-1}$) (*3*).

In the oxidized Glu47Ala mutant (Fig. 1C), the absorption band specific for center II was even more blue-shifted ($\lambda_{max}$ = 580 nm). The value of the molar extinction coefficient at 580 nm was 1,600 mM$^{-1}$ cm$^{-1}$, again slightly lower than that for the wild-type SOR. With both mutants, comparable spectra were obtained when O$_2^-$ was provided by the xanthine/xanthine oxidase system (data not shown).

Oxidation of wild-type and mutant SORs by a slight molar excess of $H_2O_2$ gave spectra comparable to those obtained with $O_2^{\cdot-}$ as an oxidant (data not shown). Addition of sodium ascorbate in excess to the $O_2^{\cdot-}$-fully oxidized Glu47Ala and Lys48Ile mutants restored the visible spectra of the as-isolated proteins (data not shown).

*EPR Spectroscopy Analysis*. The 4 K EPR spectra of Glu47Ala and Lys48Ile mutants displayed resonances at g = 7.7, 5.7, 4.1 and 1.8, similar to those obtained for the wild-type SOR (*3*) (data not shown). These resonances are typical of those for a distorted $FeS_4$ center and are assigned to center I (*5, 7*). The 4 K EPR spectrum of the $O_2^{\cdot-}$-oxidized Lys48Ile mutant presented an additional signal at g = 4.3 (data not shown), as also observed for wild-type SOR (*3*), attributed to the oxidized center II (*7*). The 4 K EPR spectrum of the oxidized Glu47Ala mutant also displayed a signal at g = 4.3 corresponding to oxidized center II. However, this signal was more complex with an additional feature at g = 4.2 (data not shown).

*Steady-state kinetics of the reaction of the Glu47Ala and Lys48Ile mutants with $O_2^{\cdot-}$*. The effect of the Glu47Ala and Lys49Ile mutations on the rate constant for the oxidation of Dfx by $O_2^{\cdot-}$ was tested using a methodology developed earlier (*3*). The kinetics of the oxidation of the iron center by $O_2^{\cdot-}$, generated by the xanthine-xanthine oxidase system in the presence of catalase, was followed spectrophotometrically at 575 or 635 nm for the Glu47Ala and Lys48Ile mutants, respectively, in the absence or in the presence of different amounts of SOD (Fig. 2). As shown in Fig. 2A, in the case of the Glu47Ala mutant, addition of large amounts of SOD resulted in the decrease of the initial rate of oxidation. The inset shows a linear plot of the reciprocal of the initial rate of oxidation of the iron center ($v_{ox}$) as a function of SOD concentration. From this plot, the concentration of SOD that decreased by 50% the rate of the reaction was determined to be 13±3 µM. The second order rate constant of the oxidation of the iron center by $O_2^{\cdot-}$ was calculated using Eq. 1 (Materials and Methods) and a value of $(1.4\pm0.4)\times10^9$ $M^{-1}$ $s^{-1}$ was obtained. This value is close to that determined for the wild-type protein under the same experimental conditions, $(0.8\pm0.2)\times10^9$ $M^{-1}$ $s^{-1}$ (data not shown and (*3*)), suggesting that the Glu47Ala mutation has almost no effect on the reaction of SOR with $O_2^{\cdot-}$.

In Fig. 2B is shown the kinetics of the oxidation of the Lys48Ile mutant by $O_2^{\cdot-}$ and the effect of increasing amounts of SOD. The reaction goes much slower and the concentration of SOD that decreased by 50% the rate of oxidation of the iron center was determined to be 0.38±0.04 µM. The second order rate constant of the oxidation of the iron center by $O_2^{\cdot-}$ was calculated to be $(4.2\pm0.5)\times10^7$ $M^{-1}$ $s^{-1}$. This value is about 20-fold lower than that determined in the case of the wild-type protein (see above), clearly showing that the Lys residue at position 48 plays an important role in the oxidation of center II by $O_2^{\cdot-}$.

*Complementation of E. coli SOD-deficient strains by Glu47Ala and Lys48Ile mutant forms of SOR*. The ability of the SOR mutated proteins to complement *E. coli* SOD deficiency was tested as described in (*13*). The *E. coli sodA sodB recA* mutant cannot grow in the presence of oxygen because of the combined lack of superoxide dismutase activity (*sodA sodB*) and the DNA strand break repair activity (*recA*) that results in lethal DNA oxidative damage (*22, 23*). As shown in Table I, the production of the Glu47Ala mutated SOR protein restored aerobic growth of *sodA sodB recA* mutant with nearly the same efficiency as that of the wild-type SOR protein. In contrast, production of the Lys48Ile SOR mutated protein did not permit colony formation after 24 h. However colonies appeared after further incubation, reaching normal size only after 72 h. Furthermore, less than 20% of anaerobically growing bacteria producing the mutated SOR were able to grow aerobically, compared to 70% of those producing wild type SOR. This indicates that production of Lys48Ile SOR can compensate for the SOD deficiency, but with much less efficiency than wild-type and Glu 47Ala SOR forms.

*Pulse radiolysis experiments and observation of reaction intermediates*.

In order to study the reaction of SOR wild-type or mutant proteins with $O_2^{.-}$ in the µs-ms time range and to detect reactive intermediate species, we have carried out pulse radiolysis experiments. Under our experimental conditions, proteins were present in large excess with regard to superoxide ([protein] = 70-150 μM, [$O_2^-$] = 3 μM).

*i) Reaction of the wild-type SOR with superoxide.* Absorbances between 450 and 700 nm increased and reached a maximum ca. 50 μs after the pulse. The absorbances then remained stable for at least 300 μs. Kinetics were analyzed at different wavelengths and were all found qualitatively identical. Fig. 3 shows two representative traces, at 550 and 630 nm. A second-order rate constant for the reaction of $(1.1\pm0.3)\times10^9$ $M^{-1}$ $s^{-1}$ was determined. These data are consistent with the very fast formation of a first reaction intermediate during the first 50 μs of the reaction. The spectrum of this intermediate could be reconstructed from the absorbances at different wavelengths obtained after 50 μs reaction (Fig. 4). This intermediate species exhibits a maximum at 610 nm and a shoulder at 550 nm (Fig. 4). The molar extinction coefficient at 610 nm for this first intermediate, is equal to 4700 $M^{-1}$ $cm^{-1}$, assuming that the reaction of SOR with 3 μM of $O_2^-$ is quantitative.

This first intermediate underwent transformations on the millisecond time scale. Fig. 5 shows two representative traces (up to 15 ms) at 580 and 630 nm. All traces between 450 and 700 nm can be described as the sum of two exponential processes, both independent of the protein concentration at all wavelengths investigated. These processes have rate constants of 500±50 $s^{-1}$ and 25±5 $s^{-1}$, respectively. Unfortunately, the final absorbances could not be determined because of the lamp instability after about 0.1 s and the reaction could not be further investigated after about 35 ms reaction time.

The spectrum of the second intermediate species formed at a rate of 500 s$^{-1}$ at its maximum concentration is shown on Fig. 5. It exhibited a spectrum with a sharp band centered at 630 nm (Fig. 4). When compared with the previous intermediate formed after 50 µs reaction (Fig. 4), the shoulder at 550 nm has vanished and the absorbance at 630 nm has increased. The extinction coefficients of the second intermediate were calculated from the fitting values, since deconvolution of the two first-order kinetic processes was necessary. A value of 6500 M$^{-1}$ cm$^{-1}$ at 630 nm was calculated.

At the end of the reaction, the difference spectrum (pulse radiolysis-treated SOR minus initial SOR) indicated that about 6 µM SOR iron center II had been oxidized (data not shown). This is likely to be the result of an initial oxidation of SOR with 3 µM of $O_2^-$ generated by pulse radiolysis, followed by a further oxidation with 3.35 µM of $H_2O_2$. 3 µM $H_2O_2$ comes from the reaction of reduction of $O_2^-$ and 0.35 µM from the radiolytic process (G($H_2O_2$)=0.7µmolJ$^{-1}$).

*ii) Reaction of the Glu47Ala mutant SOR with superoxide.* Absorbance changes reached a plateau 50 µs after the pulse, at all wavelengths between 450 and 700 nm (data not shown), as observed in the case of the wild-type SOR (Fig. 3). It displays a peak at 630 nm and a broad shoulder at around 580 nm, comparable to that of the first intermediate during the reaction of wild-type SOR. The rate constant for the formation of this first intermediate is equal to $(1.2\pm0.2)\times10^9$ $M^{-1}$ $s^{-1}$. This value is identical to that found for the wild-type protein. The extinction coefficient of this species at 630 nm was found to be 5830 $M^{-1}$ $cm^{-1}$, assuming a quantitative reaction of SOR with 3 µM $O_2^{.-}$.

On a longer time scale, absorbance changes could be fitted with two successive first order reactions, at all the wavelengths investigated between 500 and 680 nm (data not shown). The observed rate constants for both reactions did not vary with the protein concentration and values of 440±50 $s^{-1}$ and 20±5 $s^{-1}$, respectively, were determined (data not shown). These rate constants are comparable to those determined for the wild-type protein. The spectrum of a second intermediate reached a maximal intensity 5 ms after the pulse (Fig. 6) and was found roughly comparable to the spectrum of the second intermediate in the case of the wild-type SOR. However, for the former, a smaller extinction coefficient value for the peak at 630 nm and a more visible shoulder at 570 nm were observed.

*iii) Reaction of the mutant Lys48Ile mutant SOR with superoxide*. In the case of the Lys48Ile mutant, at all the wavelengths observed between 475 and 700 nm, the initial absorbance changes occurred much more slowly. Fig. 7 shows a representative trace of the time-dependent absorbance change at 600 nm. All kinetic traces between 475 and 700 nm could be described by the sum of two exponential processes, with a maximum of absorbances at all wavelengths obtained 1 ms after the pulse. The rate of formation of this first 1 ms intermediate is equal to $(3.8\pm0.6)\times10^7$ $M^{-1}$ $s^{-1}$. This value is more than one order of magnitude lower than that determined in the case of the wild-type and the Glu47Ala mutant. The reconstituted spectrum of this intermediate shows one broad band centered at 600 nm without any shoulder (Fig. 8). It differs clearly from that found for the wild-type SOR (Fig. 4). This species then underwent a transformation, which seems to be complete about 10 ms after the pulse. The rate constant of this transformation did not depend on the protein concentration and was found to be $300\pm30$ $s^{-1}$. The reconstituted absorption spectrum, 10 ms after the pulse, is shown on Fig. 8. It is comparable to the spectrum of the 1 ms intermediate, albeit with a smaller intensity and a significant red shift from 600 nm to about 615 nm for the $\lambda_{max}$. It is followed by a much slower transformation which however is difficult to follow with our device (data not shown). The final absorbances were then measured on a UV-visible spectrophotometer, when the reaction was completed. The difference spectrum (pulse radiolysis treated SOR minus initial SOR) was consistent with the oxidation of the iron center II with $O_2^{\cdot-}$ generated by pulse radiolysis, plus an oxidation with $H_2O_2$, coming from superoxide reduction and from its radiolytic formation (data not shown).

DISCUSSION

In the present work, we have investigated the reaction of SOR from *D. baarsii* with superoxide by pulse radiolysis experiments. Pulse radiolysis is one of the rare techniques that allow instantaneous and specific production of superoxide, in a dose-controlled manner, and is then particularly appropriate to study the reaction of SOR. The reaction was followed spectrophotometrically in the 450-750 nm region, since the active site of SOR, the iron center II, exhibits one absorption band at 644 nm in its oxidized redox state, assigned to a cysteinyl ligand to iron (III) charge transfer transition (*7*).

The data presented here are consistent with a very fast bi-molecular reaction of iron center II with superoxide, with a second order rate constant of about 1 x $10^9$ $M^{-1}$ $s^{-1}$, followed by the formation of two successive transient intermediate species, with spectral absorbances in the 500-750 nm range. The rate constants for the formation and disappearance of the two intermediates are summarized in Table II. Unfortunately, because of the lamp instability of the apparatus, we were not able to investigate the last part of the reaction, which gives the final product of the reaction, i.e. the oxidized iron center II of SOR and $H_2O_2$. On the basis of the rates of appearance and decay of the observed transient species and of their independence on protein concentration, we conclude that these species are true active site intermediates formed during the reaction of SOR with $O_2^-$.

Although the UV-visible absorption properties cannot provide enough information in order to identify the observed intermediates, we make the proposal that these transient species are the corresponding iron (II)-superoxide and iron (III)-peroxide species, as illustrated in Scheme 2. The first intermediate is suggested to be an iron (II)-superoxide species, resulting from the very fast binding of $O_2^-$ to the solvent accessible coordination site of the reduced iron center II (*9, 16*). The second intermediate is proposed to derive from an electron transfer from the iron to $O_2^-$ to give a species which would probably be an iron (III)-peroxide complex. Nonheme mononuclear peroxo-iron(III) species, which have been characterized in the case of several model compounds, are known to exhibit electronic transitions in the 500-600 nm region, associated with the peroxo to the iron (III) charge transfer transition (*24, 25, 26*). Finally, the last slowest part of the reaction, which is partly lacking from our analysis, would correspond to the release of $H_2O_2$. Further studies are required to identify the steps where protonation of the peroxide takes place.

In order to further characterize these intermediate species, clearly more investigations are required. However, because of the very rapid kinetics of formation and disappearance of the first intermediate, it might be difficult or even impossible to accumulate and stabilize it in order to apply other spectroscopic techniques for more detailed characterization. In the case of the second intermediate, rapid freeze quench techniques, coupled to EPR or Raman spectroscopies may give valuable information on the nature of this species.

In previous work and here, we have shown that the rate constant for the oxidation of center II by $O_2^-$, determined by steady-state measurements, is almost identical to the value obtained by pulse radiolysis for the formation of the first intermediate species (Table II). Although the overall process of the oxidation of center II by $O_2^-$ (rate-limited at least by the disappearance of the second intermediate at 25 s$^{-1}$) is much slower than the initial bimolecular process, the steady-state experiments are in very good agreement with the pulse radiolysis data. As a matter of fact, during the steady-state kinetics, competition with SOD allows to determine a rate constant only for the direct bimolecular reaction between $O_2^-$ and the iron center II. No kinetic information for the following intramolecular steps that will produce the oxidized form of iron center II could be deduced from these experiments.

In order to get deeper insight into the mechanism of SOR, we have investigated the role of two amino-acid residues, Glu47 and Lys48. The two available three-dimensional structures of SORs from *Desulfovibrio desulfuricans* (*9*) and *Pyrococcus furiosus* (*16*) show that the two residues homologous to Glu47 and Lys48 in the case of *D. baarsii* are the charged residues closest to the sixth free coordination position of the iron center II and thus are potentially important for catalysis (Scheme 1). In order to evaluate the importance of these residues, we have changed Glu47 and Lys48 into Ala and Ile, respectively. The purified mutant proteins were correctly folded and fully metallated. They were both reactive towards superoxide, allowing comparison to the wild-type SOR.

The reactivity of the Glu47Ala mutant with regard to superoxide is similar to that of the wild-type protein. As shown by pulse radiolysis experiments, the two observed intermediate species were characterized with formation rate constants and visible spectra comparable to those observed in the case of the wild-type SOR (Table II). This suggests that Glu47 does not participate to the first steps of the reaction. Accordingly, from the three-dimensional structure of *P. furiosus*, it was suggested that Glu14, corresponding to Glu47 in *D. baarsii*, did not bind to the reduced iron center II (*16*) (Scheme 1). Thus, in both wild-type and mutant SORs, a coordination position is available. The only difference between the mutant and wild-type SOR resides in the spectroscopic properties of the final product. The absorption band of center II of the oxidized mutant is blue-shifted with respect to wild-type SOR, from 644 to 580 nm, and the 4 K EPR spectra display slight differences. The three-dimensional structure of SOR from *P. furiosus* also suggested that in the oxidized state, Glu14 occupies the sixth coordination site of the iron center (*16*). It is thus not surprising that removal of Glu47 would result in significant modifications of the electronic properties and geometry of the oxidized iron center II.

Finally, in agreement with the absence of effects of the mutation on the reactivity of iron center II towards superoxide, we have found that expression of the *sor* Glu47Ala mutant gene restores the capability for aerobic growth of an oxygen-sensitive *sodA sodB recA E. coli* mutant strain as efficiently as does the wild-type *sor* gene. Thus the mutant gene still provides strong protective effects against oxidation stress. In conclusion, Glu47 does not appear to be absolutely required for a functional SOR within the cell.

During submission of this work, Coulter et al. reported a similar investigation on the SOR from *Desulfovibrio vulgaris* and made the appealing suggestion that Glu47 might nevertheless participate to the release of the peroxide by protonation or nucleophilic substitution (*27*).

The Lys48Ile SOR mutant exhibits a 20-fold smaller second order rate constant for the reaction with superoxide with regard to the wild-type protein when measured by steady-state kinetics (Table II). During pulse radiolysis experiments, two intermediates species were also observed, as in the case of the wild-type SOR. However, the second order rate constant for the formation of the first species is 30 times lower than the corresponding value found for the wild-type protein (Table II). This is in line with the value determined from the steady-state kinetics. The reconstituted spectrum of the first intermediate of the Lys48Ile mutant is slightly different from that of the wild-type protein. On the other hand, the kinetics of the formation of the second intermediate in the case of the Lys48Ile mutant was comparable to that found with the wild-type SOR (Table II). The spectrum of the second intermediate is similar to that of the wild-type, except for a smaller extinction coefficient of the absorption band. Once again, the following steps of the reaction ending up with the formation of the reaction products, $H_2O_2$ and iron (III), could not be observed by our techniques, and possible effects of the Lys mutation on these steps could not be investigated.

From the data presented here, we can conclude that the Lys48 mutation greatly affects the rate of the direct reaction of $O_2^-$ with the iron center II and that of the formation of the first intermediate, without notably disturbing the rate of formation of the second intermediate. It is tempting to suggest that Lys48, the positively charged residue closest to the iron center II, plays a role in directing and stabilizing $O_2^-$ to the active site, and is then required to allow the very fast reaction of SOR with superoxide. The physiological importance of Lys48 was confirmed from the observation that the Lys48Ile *sor* mutant gene has a very poor antioxidant protective effect on the *sodA sodB recA E. coli* mutant strain, as compared to wild-type and Glu47Ala *sor* genes. A similar phenotype was indeed observed when the wild-type *sor* gene was induced by very low levels of IPTG, resulting in a low production of SOR wild-type protein ((*13*) and unpublished data).

Superoxide reductases are exquisite systems for understanding how biological iron centers deal with superoxide at the molecular level. As a matter of fact, they are much simpler than superoxide dismutases as $O_2^-$ is implicated in only one step, oxidation of ferrous iron. The work reported here, with the original observation of reaction intermediates and the investigation of the first site-directed mutants, provide the foundation of further investigations for a better understanding of the chemistry of this new class of antioxidant enzymes.

ACKNOWLEDGEMENTS

We are grateful to Dr. Eric Forest for mass spectrometry experiments, to Mathilde Louwagie for N-terminal amino-acid sequence determination and to Dr. Véronique Ducros for atomic absorption spectroscopy. We acknowledge Drs. Stéphane Ménage and Jacques Gaillard for helping in EPR experiments and Chantal Falco for technical assistance.

FIGURE LEGENDS

Figure 1. Effect of $O_2^-$ on the visible absorption spectra of wild-type, Glu47Ala and Lys48Ile SOR of *D. baarsii* at 25°C. The microcuvette (100 µl final) contained 180 µM of SOR wild-type (A), Lys48Ile (B), or Glu47Ala (C), in 50 mM potassium phosphate buffer, pH 7.6, and 500 units/ml catalase. Successive additions of 180 µM $KO_2$, from a 2 mM $KO_2$ stock solution dissolved in 100% $Me_2SO$ (14M), were performed. After each addition, a spectrum was recorded. A. Wild-type SOR. From bottom to top : 0, 2 and 4 equivalents $KO_2$. B. Lys48Ile mutant. From bottom to top : 0, 3, 6, and 8 equivalents $KO_2$. C. Glu47Ala mutant. From bottom to top : 0, 2, and 4 equivalents $KO_2$. Insets show the difference spectra obtained by subtracting the initial spectrum from each spectrum after $KO_2$ addition.

Figure 2. Steady-state kinetics of oxidation of the Glu47Ala and Lys48Ile SOR mutants by $O_2^-$. Oxidation of iron center II was followed spectrophotometrically, at 25°C, by the increase of absorbance at 575 nm for Glu47Ala SOR mutant (A) and at 635 nm for Lys48Ile SOR mutant (B), in the presence of different amount of SOD. The cuvette contained (500 µl final volume) 18 µM SOR mutants, 50 mM potassium phosphate buffer pH 7.6, 800 mM hypoxanthine, 500 units catalase. The oxidation was initiated by adding

0.045 U of xanthine oxidase. A. Glu47Ala mutant with: (○) 0, (●) 4, (♦) 6, (Δ)12 μM SOD. B. Lys48Ile mutant with: (○) 0, (●) 0.1, (♦) 0.3, (Δ) 0.5 μM SOD. Insets represent the reciprocal of the initial velocity of the oxidation of the iron center ($v_{ox}$) as a function of SOD concentration.

Figure 3. Time-dependent of absorbance changes at 550 and 630 nm during the reaction of the wild-type SOR wild-type (100 μM) with $O_2^-$ (3 μM) generated by pulse radiolysis. The dashed lines are the best fit assuming a single exponential process.

Figure 4. Reconstituted spectra of the solution during the reaction of wild-type SOR (100 μM) with $O_2^-$ (3 μM) generated by pulse radiolysis, 50 μs (●) and 5 ms (❏) after the pulse.

Figure 5. Time-dependent absorbance changes at 580 and 630 nm during the reaction of the wild-type SOR (100 μM) with $O_2^-$ (3 μM) generated by pulse radiolysis. The dashed lines are the best fit assuming two exponential processes.

Figure 6. Reconstituted spectra of the solution during the reaction of Glu47Ala SOR mutant (100 μM) with $O_2^-$ (3 μM) generated by pulse radiolysis, 50 μs (●) and 5 ms (❏) after the pulse.

Figure 7. Time-dependent absorbance change at 600 nm during the reaction of the as isolated Lys48Ile SOR mutant (100 μM) with $O_2^-$ (3 μM) generated by pulse radiolysis. The dashed lines are the best fit assuming two exponential processes.

Figure 8. Reconstituted spectra of the solution during the reaction of Lys48Ile SOR mutant (100 μM) with $O_2^-$ (3 μM) generated by pulse radiolysis, 1 ms (●) and 10 ms (□) after the pulse.